\documentclass[12pt]{article}
\usepackage{times}
\usepackage{geometry}
\usepackage{floatflt}
\usepackage[utf8]{inputenc}
\usepackage{enumitem,amssymb}
\usepackage{ragged2e}
\newlist{thematic}{itemize}{8}
\setlist[thematic]{label=$\square$}
\usepackage{pifont}
\newcommand{\cmark}{\ding{51}}%
\newcommand{\done}{\rlap{$\square$}{\raisebox{2pt}{\large\hspace{1pt}\cmark}}%
\hspace{-2.5pt}}

\usepackage{jheppub}   
\usepackage[dvipsnames,svgnames,x11names]{xcolor}
\usepackage{sectsty}
\usepackage[outercaption]{sidecap}
\sidecaptionvpos{figure}{c}

\definecolor{DarkGreen}{rgb}{0.0, 0.3, 0.0}
\definecolor{purple}{rgb}{0.5, 0.0, 0.5}
\definecolor{red}{rgb}{1, 0.0, 0.0}
\definecolor{green}{rgb}{0, 1.0, 0.0}

\newcommand{\Moon}{$M_{Moon}$}                          

\newcommand{\apjl}{\rm ApJL}
\newcommand{\apjs}{\rm ApJS}

\newcommand{\aap}{\rm A$\&$A}

\def\ao{{\it Appl.~Opt.}}           

\def\3he{$^3{\rm He}$}
\def\arcdeg{\hbox{$^\circ$}}

\def\lsim{\mathrel{\lower2.5pt\vbox{\lineskip=0pt\baselineskip=0pt
           \hbox{$<$}\hbox{$\sim$}}}}

\def\gsim{\mathrel{\lower2.5pt\vbox{\lineskip=0pt\baselineskip=0pt
           \hbox{$>$}\hbox{$\sim$}}}}

\begin{document}
\huge
\begin{minipage}{\textwidth}
\raggedright
Astro2020 Science White Paper \linebreak

Unveiling the Galaxy Cluster -- Cosmic Web Connection with X-ray observations in the Next Decade
\linebreak
\end{minipage}

\normalsize
\begin{minipage}{\textwidth}
\raggedright

\noindent \textbf{Thematic Areas:} \hspace*{60pt} $\square$ Planetary Systems \hspace*{10pt} $\square$ Star and Planet Formation \hspace*{20pt}\linebreak
$\square$ Formation and Evolution of Compact Objects \hspace*{31pt} $\done$ Cosmology and Fundamental Physics \linebreak
  $\square$  Stars and Stellar Evolution \hspace*{1pt} $\square$ Resolved Stellar Populations and their Environments \hspace*{40pt} \linebreak
  $\done$    Galaxy Evolution   \hspace*{45pt} $\square$             Multi-Messenger Astronomy and Astrophysics \hspace*{65pt} \linebreak

\textbf{Principal Authors:}

Name: Stephen A. Walker$^1$, Daisuke Nagai$^2$ 	
 \linebreak						
Institution: 1) NASA GSFC, 2) Yale University,
 \linebreak
Email: stephen.a.walker@nasa.gov; daisuke.nagai@yale.edu
 \linebreak
Phone: +1 (301) 286-9882; +1 (203) 432-5370
 \linebreak
 
\textbf{Co-authors:} A. Simionescu (SRON), M. Markevitch (NASA GSFC), H. Akamatsu (SRON), M. Arnaud (CEA), C. Avestruz (U.Chicago), M. Bautz (MIT), V. Biffi (CfA), S. Borgani (UniTS/INAF), E. Bulbul (CfA), E. Churazov (MPA), K. Dolag (USM/MPA), D. Eckert (MPE), S. Ettori (INAF), Y. Fujita (Osaka), M. Gaspari (Princeton), V. Ghirardini (CfA), R. Kraft (CfA), E. T. Lau (Miami), A. Mantz (Stanford),  K. Matsushita (TUS), M. McDonald (MIT), E. Miller (MIT), T. Mroczkowski (ESO), P. Nulsen (CfA), N. Okabe (Hiroshima), N. Ota (Nara), E. Pointecouteau (IRAP), G. Pratt (CEA), K. Sato (Saitama), X. Shi (SWIFAR), G. Tremblay (CfA), M. Tremmel (Yale), F. Vazza (Bologna), I. Zhuravleva (U.Chicago), E. Zinger (Heidelberg), J. ZuHone (CfA)
\linebreak
\justify

\textbf{Abstract:} 
In recent years, the outskirts of galaxy clusters have emerged as one of the new frontiers and unique laboratories for studying the growth of large scale structure in the universe. Modern cosmological hydrodynamical simulations make firm and testable predictions of the thermodynamic and chemical evolution of the X-ray emitting intracluster medium. However, recent X-ray and Sunyaev-Zeldovich effect observations have revealed enigmatic disagreements with theoretical predictions, which have motivated deeper investigations of a plethora of astrophysical processes operating in the virialization region in the cluster outskirts. Much of the physics of cluster outskirts is fundamentally different from that of cluster cores, which has been the main focus of X-ray cluster science over the past several decades. A next-generation X-ray telescope, equipped with sub-arcsecond spatial resolution over a large field of view along with a low and stable instrumental background, is required in order to reveal the full story of the growth of galaxy clusters and the cosmic web and their applications for cosmology. 
\end{minipage}

\pagebreak

\section{Introduction}
\vspace{-2mm}

\noindent As the largest gravitationally bound structures in the universe, galaxy clusters continue to grow and accrete matter in their outskirts (see \cite{Walker2019} for a review). The majority of the gas lies in the outskirts beyond the cluster's virial radius (denoted by $r_{200}$\footnote{The radius within which the enclosed total mean overdensity is $200$ times the critical density of the Universe.}) and in the intergalactic medium (IGM) within filaments that connect clusters to the cosmic web. However, the low gas density means that these regions are extremely faint in X-rays, and at present beyond the sensitivity limits of current X-ray telescopes (Fig.\ref{LSS}, left panel). As a result, the full story of large-scale structure formation is hidden from us. Cosmological simulations of galaxy cluster formation predict the outskirts to be a hive of activity with gas continuing to accrete, and ongoing mergers with small sub-clusters and clumps of gas (Fig.\ref{LSS}, right panels). A plethora of unexplored structure formation physics is believed to be operating in the outskirts, and these physical processes are fundamentally different from the physics in the cores of clusters that has been the focus of X-ray cluster science over the past several decades. 

The outskirts of galaxy clusters is a new territory for addressing the following outstanding questions at the crossroads of cosmology and astrophysics: \textbf{How do galaxy clusters grow?} \textbf{How do they connect to the cosmic web?} \textbf{What is the chemical composition in and around the most massive structures in the Universe?} A great leap forward in the sensitivity of X-ray telescopes is required to answer these questions. 

\vspace{-4mm}
\section{How do galaxy clusters grow and connect to the cosmic web?}
\begin{figure}[t]
\hbox{
\begin{minipage}{0.51\textwidth}
\includegraphics[width=1.0\textwidth]{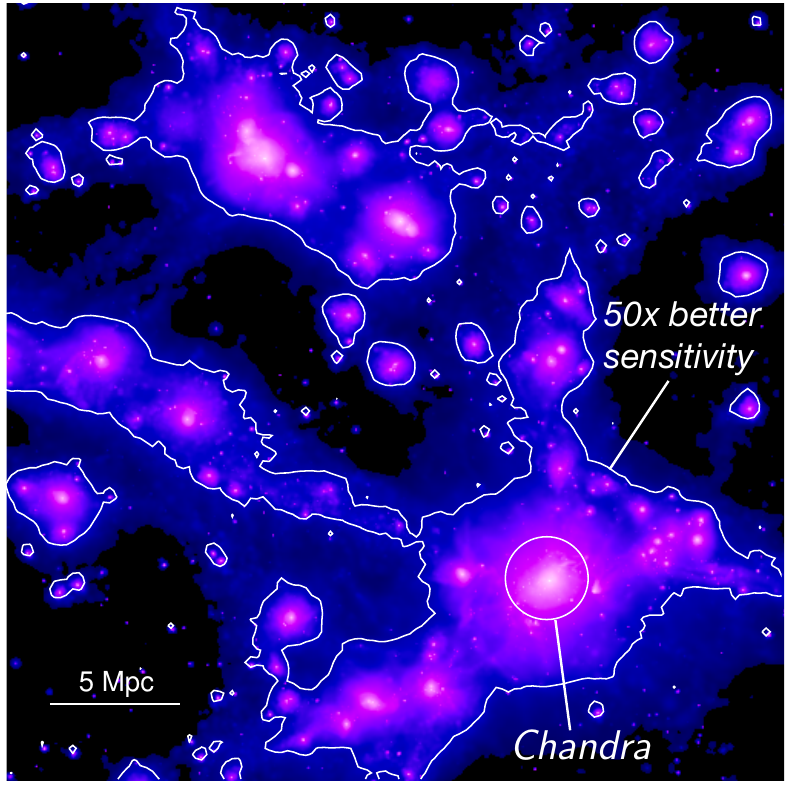}
\end{minipage}
\hfill
\hspace{0.5cm}
\begin{minipage}{0.40\textwidth}
\includegraphics[width=1.0\textwidth]{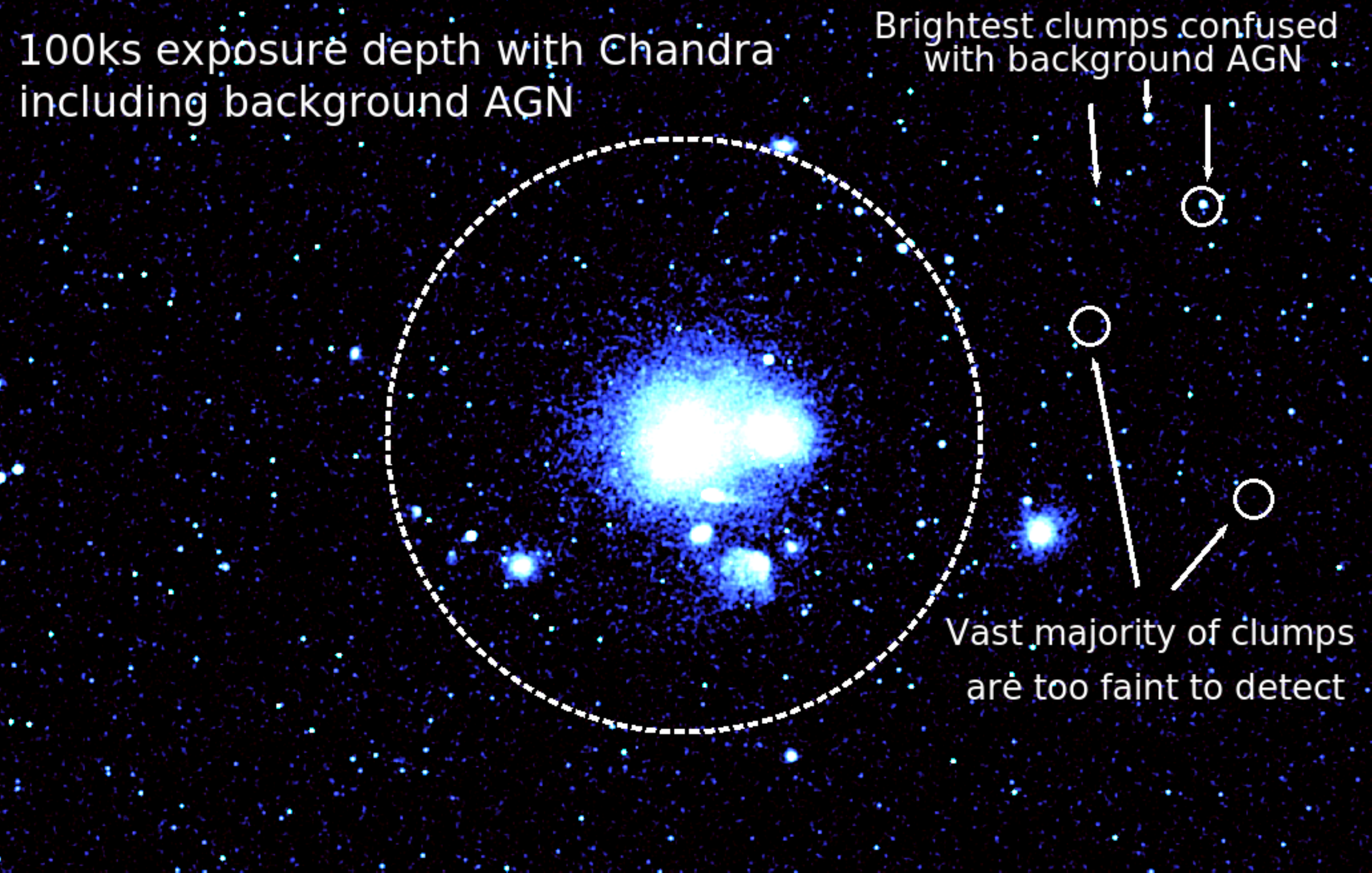}
\includegraphics[width=1.0\textwidth]{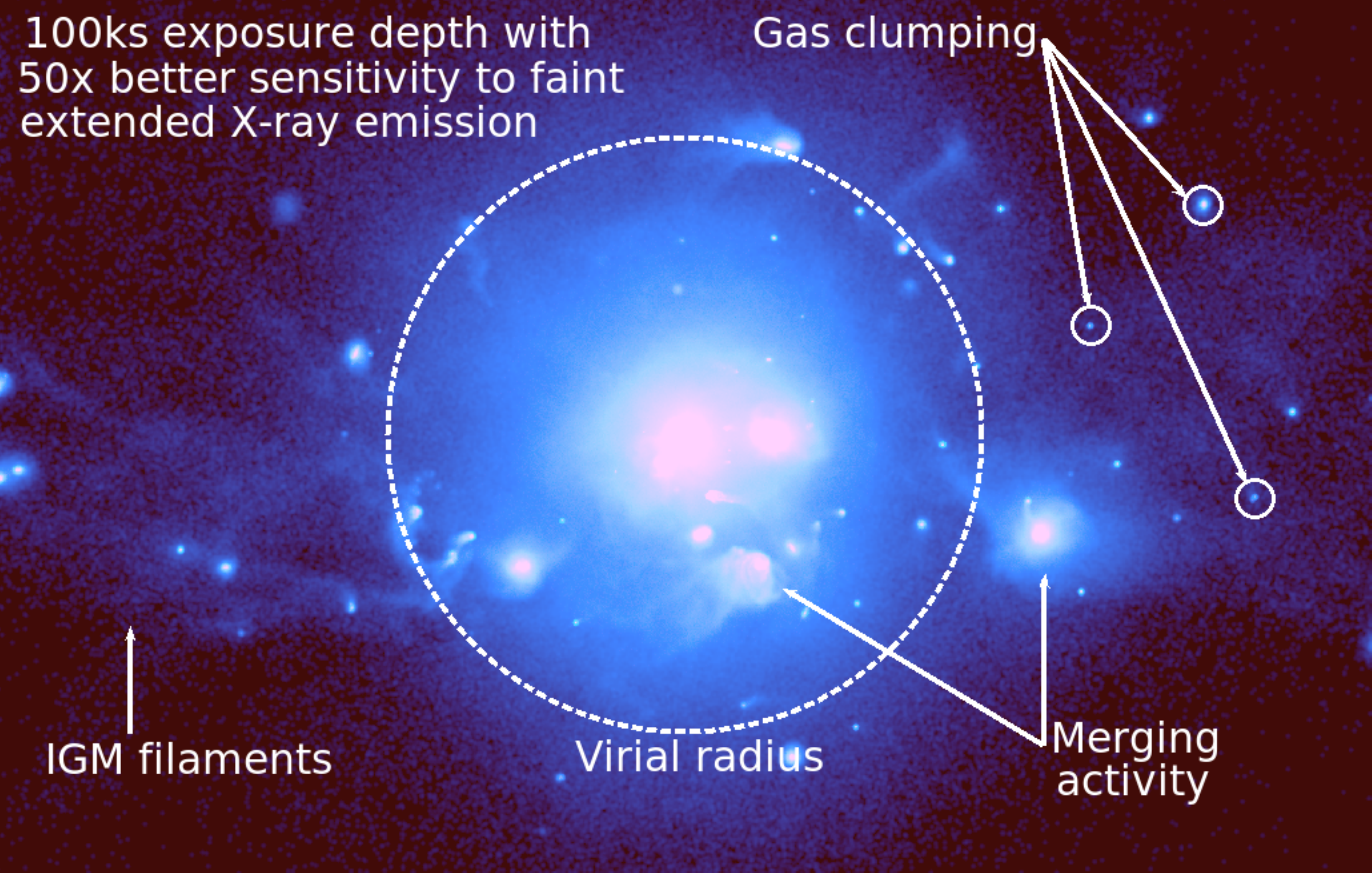}
\end{minipage}
}
\caption{{\it Left:} Hydrodynamical cosmological simulation of large-scale structure formation, showing that galaxy clusters reside at the nodes of IGM filaments \cite{Dolag2006}. With \textit{Chandra} and \textit{XMM-Newton}, only the central regions of clusters can be explored in detail (inner white circle). To begin to explore large-scale structure in its entirety (outer white contours) requires at least a factor of 50 improvement in sensitivity to low surface brightness extended emission. {\it Right:} Simulated X-ray mosaic of the RomulusC cluster \cite{Tremmel2019}, a low mass (10$^{14}$ M$_{\odot}$) cluster at $z=0.05$, comparing 100ks deep coverage from \textit{Chandra} (top) and \textit{AXIS/Lynx} (bottom). The mock \textit{Chandra} simulation includes background point sources from the cosmic X-ray background. Due to the high background and small collecting areas of current X-ray telescopes, we are only able to see the brightest X-ray emission in the cluster cores. The bulk of the ICM in the outskirts beyond the virial radius (dashed white circle), and the IGM filaments that connect clusters together, remain hidden from view.} 
\label{LSS}
\end{figure}

\begin{figure}
\hbox{
\hspace{-0.5cm}
 \includegraphics[width=0.52\textwidth]{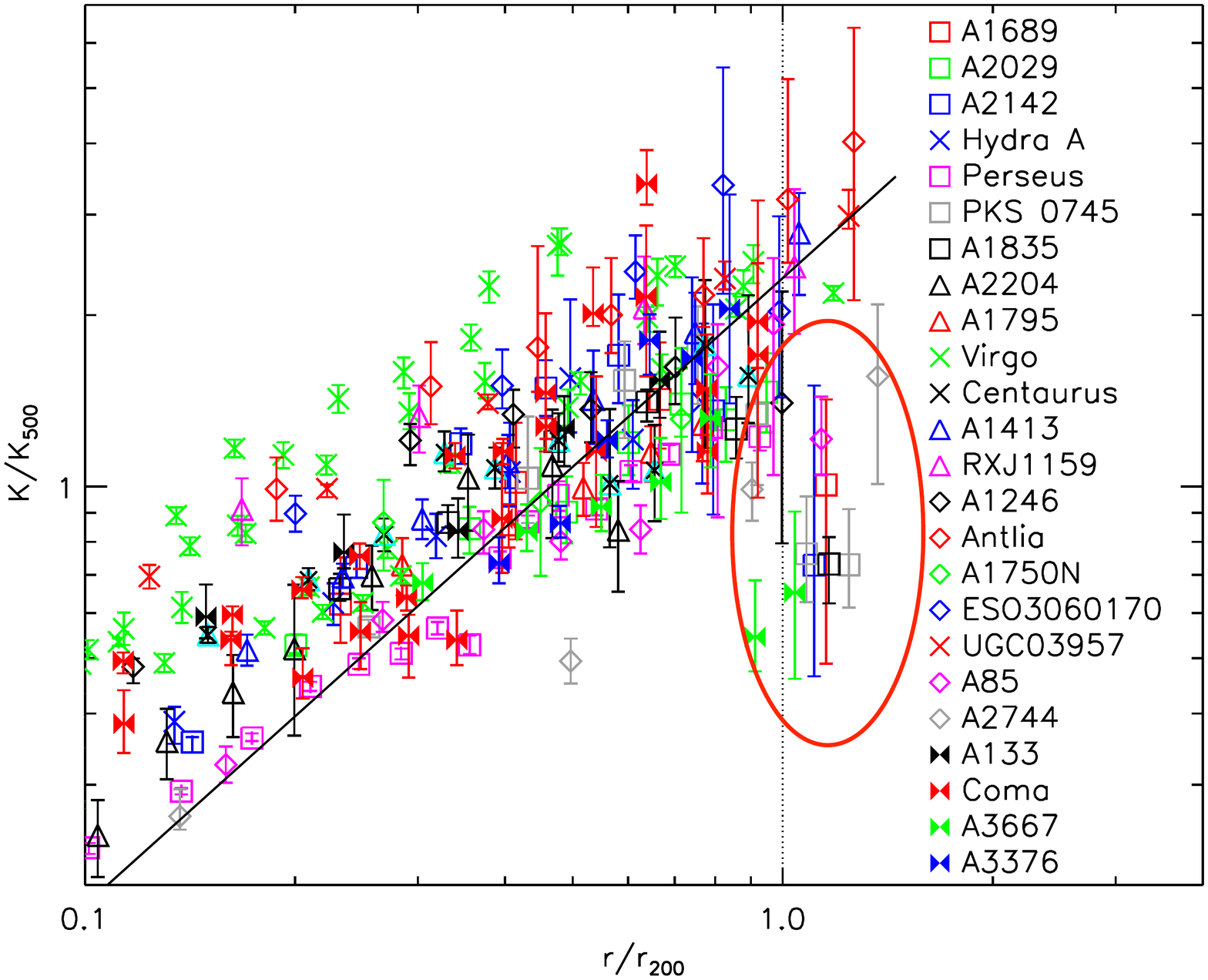}
\hspace{-0.3cm}
 \includegraphics[width=0.47\textwidth]{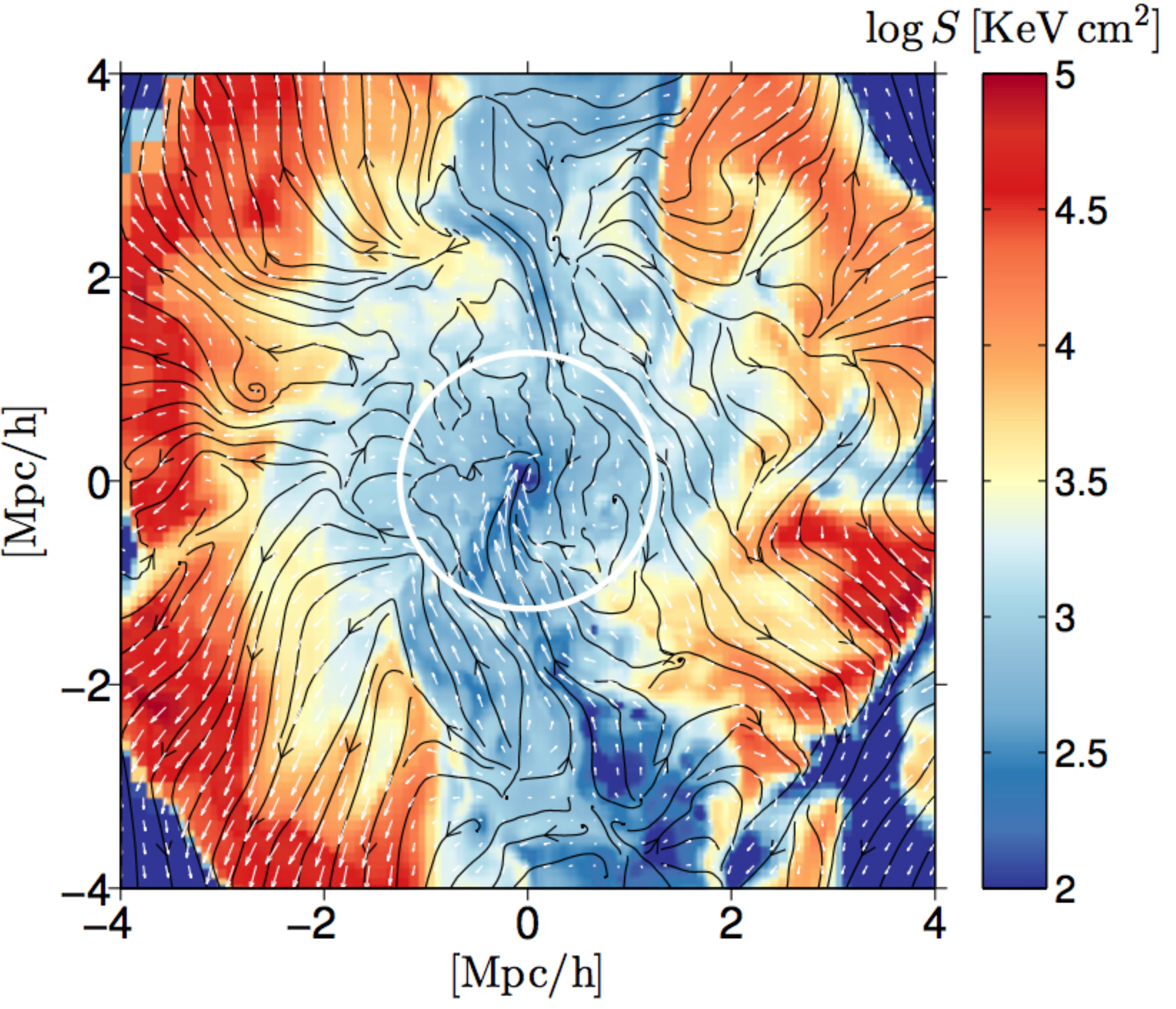}
}
\vspace{-4mm}
\caption{\emph{Left panel}: Disagreement between the observed entropy profiles in X-ray from Suzaku (coloured points) and the entropy profile predicted by gravitational collapse (solid black line, \cite{voit05}) just beyond the virial radius (highlighted by the red oval) \cite{Walker2019}. 
\emph{Right panel}: Simulated entropy map of a galaxy cluster, illustrating filamentary streams of high-density, low-entropy gas penetrating deep into the cluster interior \cite{Zinger2016}.}
\vspace{-3mm}
\label{fig:entropy}
\end{figure}

\vspace{-2mm}
Our current understanding of the gas in cluster outskirts has been achieved through recent pioneering measurements of the intracluster medium (ICM) in X-rays and microwaves via the Sunyaev-Zeldovich (SZ) effect that map the hot gas distribution near the virial radius of nearby galaxy clusters. The {\em Suzaku} X-ray Observatory, with its low and stable background afforded by its low Earth orbit, was able to obtain temperature and density measurements out to the virial radius \cite[e.g.,][]{simionescu11,Walker13_1}. Combining density measurements from deep \textit{XMM-Newton} observations of nearby bright clusters with \textit{Planck} SZ pressure profile data has also allowed statistical constraints on the thermodynamic profiles of the outskirts ICM to be made \cite{Ghirardini18b,Eckert18}.
Initial results from these observations have been puzzling, as they disagree with theoretical predictions \cite[e.g.,][see also Fig.\ref{fig:entropy}, left panel]{simionescu11,Walker13_1,Ghirardini18b,Eckert18}. In particular, the gas entropy just outside the virial radii has been observed to lie well below the predictions from purely gravitational collapse (shown as black line in the left panel of Fig.\ref{fig:entropy}).
Future high angular resolution and large field of view (FoV) X-ray and SZ observations with high sensitivity are required to address the apparent disagreement, and to capture the rich structures in cluster outskirts predicted by modern cosmological hydrodynamical simulations (see Fig.\ref{LSS}, right panels).  This will allow us to investigate the physics of galaxy and cluster formation, especially in the unexplored territory where matter from the cosmic web accrete into galaxy clusters. Below, we highlight several transformative cluster outskirt sciences for the coming decade. 

\medskip
\noindent
{\bf ICM Clumping:} 
Both \textit{Suzaku} and \textit{XMM-Newton/Planck} measurements of the ICM outskirts have found evidence for increasing levels of gas clumping, as predicted by cosmological simulations \citep{nagai11,zhuravleva13,roncarelli13,vazza13,rasia14,battaglia15,Planelles17}. Gas clumping is believed to be due to infalling gas substructures.  Simulations indicate that the level of gas clumping should increase dramatically in the (1-2)$r_{200}$ region of clusters, where the infalling clumps have not yet been entirely ram-pressure stripped by the ICM. If not resolved, clumping will bias gas profile measurements by overestimating the gas density and gas mass fraction, and underestimating the gas temperature, leading to biases in hydrostatic mass estimates. Correcting for clumping will help explain the lower observed entropy compared to predictions, the excess baryon fraction beyond the cosmic value, and help correct biases in hydrostatic mass. Detecting and resolving gas clumps, that are associated with infalling galaxies in the outskirts of clusters, is also crucial for gaining insights into the stripping of their circumgalactic medium (CGM) through their interactions with the ICM, quenching of cluster galaxies, and chemical enrichment process in the outskirts of galaxy clusters (see \S\ref{sec:metals}). Current X-ray telescopes are unable to resolve faint gas clumps that contribute to the bulk of clumping (Fig. \ref{LSS}, right), and their physical properties are almost completely unexplored. Future X-ray observations with high sensitivity and high spatial resolution, such as \textit{AXIS/Lynx}, will allow routine detection of clumps
and open up a new window for studying the physical processes of galaxy and cluster formation and evolution.

\medskip
\noindent
{\bf Bulk \& Turbulent Gas Motions:} 
Gas motions in cluster outskirts are generated from mergers and accretion during cluster formation. Thus the amount of these gas motions is a direct probe of the cluster's dynamical state.  Cosmological hydrodynamical simulations predict that such non-thermal pressure becomes more significant with radius in the density stratified ICM, reaching $\sim 50\%$ of the thermal gas pressure at $r\approx 1.5r_{200}$ \citep{nelson14,shi15}. Such non-thermal pressure is believed to be responsible for biasing hydrostatic cluster masses at a level of $10\%-30\%$ \citep{lau09, lau13, Biffi2016, shi16d, shi18, Ettori2019} by providing extra support against gravity. Upcoming high-resolution spectral X-ray observations with \textit{Athena} and \textit{Lynx} will enable us to directly measure the level of bulk and turbulent gas motions in cluster outskirts through Doppler shifting and broadening of the X-ray emission lines from the ICM \cite[][see also the white paper by Bulbul et al.]{Ota2018,Roncarelli2018}, allowing us to fully account for the hydrostatic mass bias when combined with gravitational lensing masses from optical surveys such as \textit{WFIRST} and \textit{Euclid} \citep{Pratt2019}. High-resolution X-ray and SZ spectral imaging observations will also provide complementary constraints on the level of non-thermal pressure support \cite{Khatri2016} and ICM microphysics \cite{Gaspari2013} in the outskirts of galaxy clusters.

\medskip
\noindent
{\bf Filamentary gas streams:}
Cosmological simulations predict that high density, low entropy gas in the cosmic web filaments can penetrate deep into the cluster interior \citep[][see Fig.\ref{fig:entropy}]{Zinger2016}. These filamentary streams transport the warm-hot gas from the outskirts to the cluster core regions, stirring up the gas, generating bulk and turbulent gas motions, and producing shocks and contact discontinuities generated by the interaction of the cold penetrating stream with the surrounding hot ICM \citep{Zinger2018}. These streams can deposit energy, mass, and enrich the ICM as they break up and mix with the surrounding gas via fluid instabilities. Direct observations of these streams are not possible at present as these low entropy, low temperature streams are predicted to be very X-ray faint. High spatial resolution X-ray observations with \textit{AXIS/Lynx}, with large azimuthal and radial coverage (from the cluster centers to large radii) will enable us to identify the transformation from cosmic web gas filaments in the outskirts into penetrating gas streams in the cluster interior, providing direct evidence for cosmic web filaments feeding the growth of galaxy clusters. 

\medskip
\noindent
{\bf Thermodynamics of accretion shocks:} 
The outer boundary of the ICM is defined by the accretion shock. This is an important location where cool-warm gas accreting from the cosmic web is shock heated to very high temperature, generating entropy via high Mach number ($M\approx 10-100$) accretion shocks \citep{molnar09,vazza09,lau15}. Detecting accretion shocks around clusters and measuring their strength will provide the first direct evidence of the primary physical process that defines the fate of the hot, diffuse baryons that make up more than half of the normal matter in the local Universe. 

\medskip
\noindent
{\bf Non-equilibrium electrons:} 
In the low-density region in the outskirts of galaxy clusters, the collision rate of electrons and protons becomes longer than the age of the universe, potentially causing the electron temperature to be lower than the ion temperature, especially in more massive and less relaxed systems \citep{rudd09,avestruz15}.\footnote{When an electron-ion plasma passes through the accretion shock, most of the kinetic energy goes into heating heavier ions, causing $T_{i} \gg T_{e}$. After the shock, electrons and ions equilibrate via Coulomb interactions over an electron-ion equilibration timescale, $t_{ei}$.} X-ray and SZ measurements of galaxy cluster outskirts, which probe the electron densities and temperatures of the ICM, therefore hold promise for shedding new light on the thermalization process operating during the cluster formation, and probing for the first time a regime where the physics of non-equilibrium plasmas beyond the simple hydrodynamics approximation becomes important in the ICM.

\vspace{-4mm}
\section{How do metals spread in galaxy clusters \& the cosmic web?}
\label{sec:metals}
\vspace{-2mm}

\begin{figure}
 \hbox{
  \includegraphics[width=0.7\textwidth]{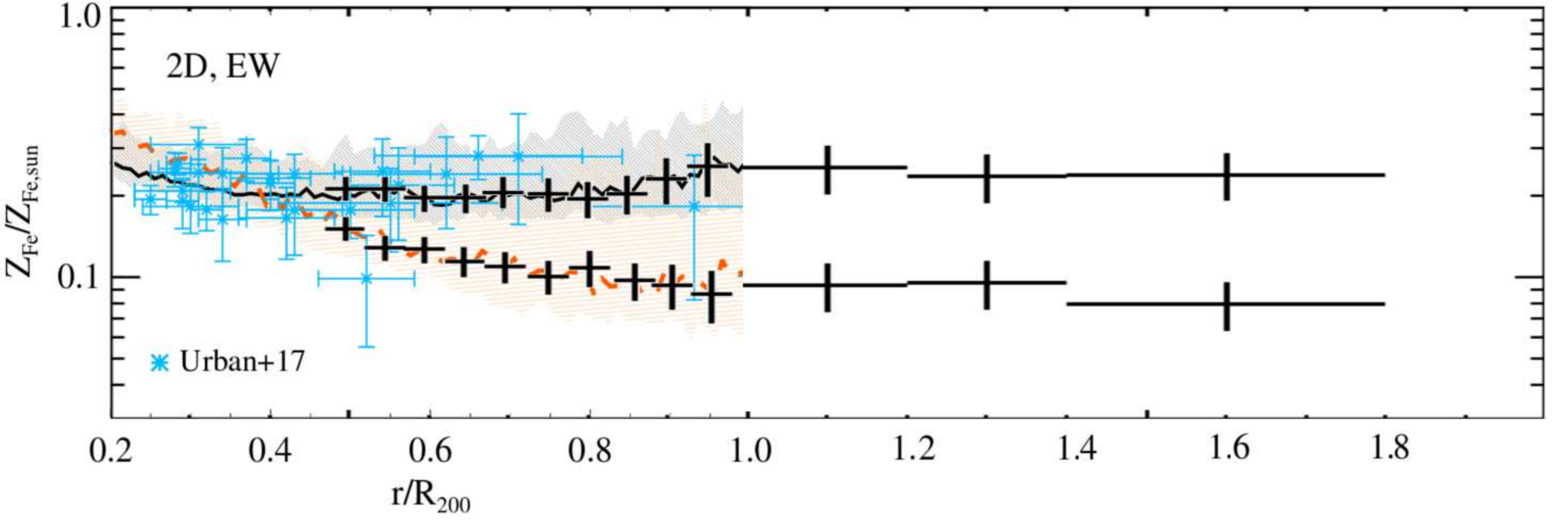}
  \hspace{0.5cm}
  \includegraphics[width=0.18\textwidth]{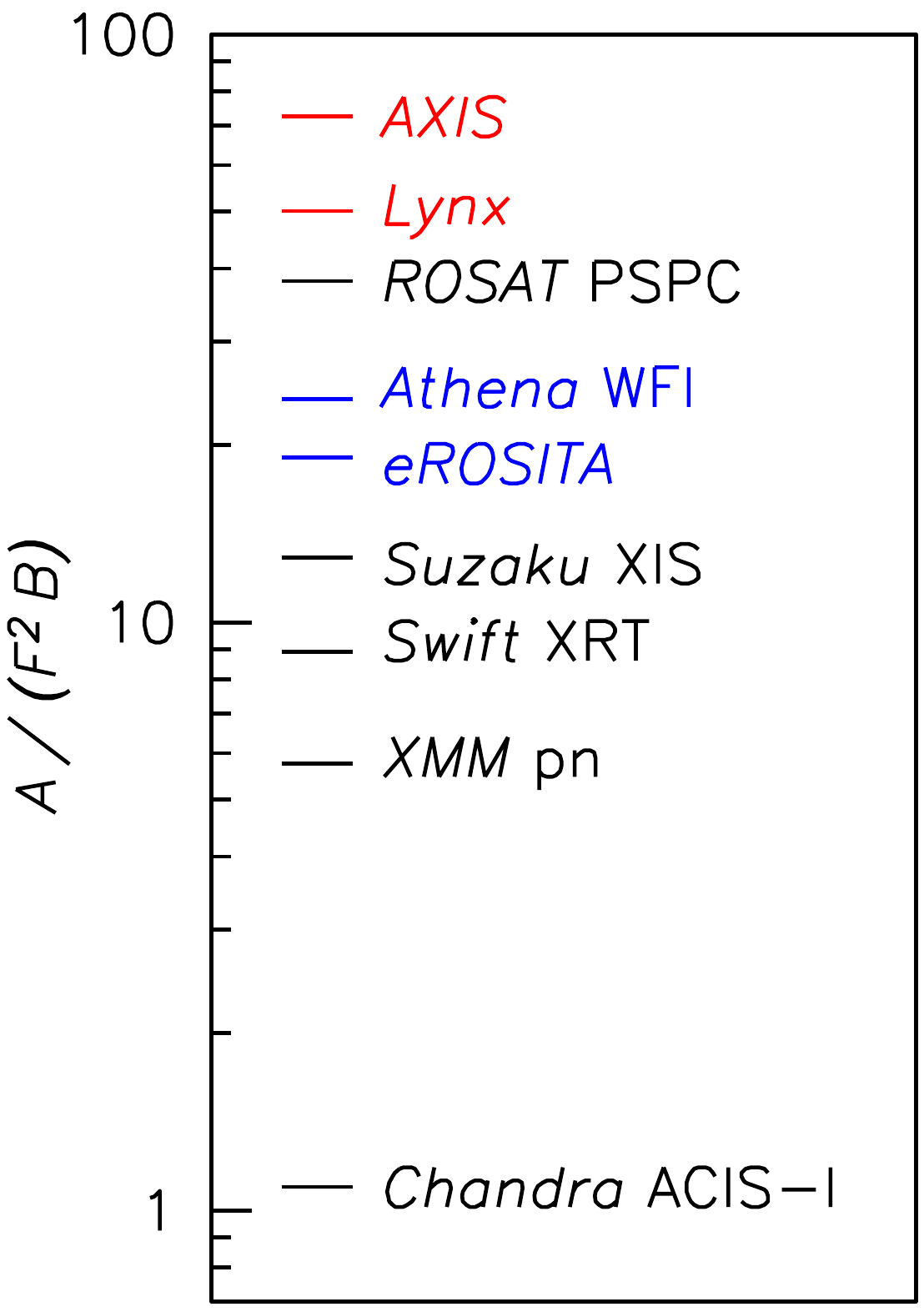}
 }
 \vspace{-0.2cm}
\caption{\emph{Left}: Simulated metallicity profiles beyond the virial radius with stellar and AGN feedback (black solid curve) and only stellar feedback without AGN (red dashed curve), highlighting AGN feedback plays an important role in elevating the metallicity level in clusters outskirts \cite{Biffi2018}. Suzaku X-ray measurements are shown in blue. Simulated data from \textit{AXIS/Lynx} are shown as the black crosses for high and low metallicity scenarios, showing that highly accurate metallicity measurements can be made out to 1.8$r_{200}$. \emph{Right}: The ratio of sky signal to detector background, which determines the sensitivity to low surface brightness emission, for major X-ray imaging instruments (blue and red are future and proposed, respectively). $A$\/ is effective area at 1 keV, $B$\/ is detector background (assuming 1 at low-Earth orbit and 4 at high-Earth orbit for simplicity) and $F$\/ is the focal length. \textit{Lynx} and \textit{AXIS} are $>$50 times more sensitive than \textit{Chandra}. 
}
\vspace{0.0cm}
\label{metals}
\end{figure}

Galaxy clusters, with their deep potential wells, are `closed-box' systems ideal for studying galaxy formation physics. In particular, metallicity in the outskirts of galaxy clusters is a powerful probe of feedback physics \cite{Werner2013, Biffi2018, Mernier2018}, as the metal distribution in the ICM is strongly dependent on the chemical enrichment histories. Feedback from active galactic nuclei (AGN) at early times ($z\gtrsim 2$) is found to be effective at removing pre-enriched gas from galaxies, spreading metals uniformly throughout the cluster outskirts (see the solid black curve in Fig.\ref{metals}, left). On the other hand, late-time enrichment leads to inhomogeneous distributions of ICM metals that rapidly decline with cluster-centric radius (red curve in Fig.\ref{metals}, left). The relative composition of various elements of the ICM originating from different types of supernova explosions (core-collapse versus Ia) also place constraints on the star formation histories and the chemical evolution of the universe \citep{Simionescu2015}. 

Measurements of metal abundance in the outskirts are extremely challenging, and are currently limited to within half the virial radius for \textit{Chandra} and \textit{XMM-Newton}. With \textit{Suzaku} only very few clusters have measurements reaching the virial radius (blue crosses in Fig.\ref{metals} from \cite{Urban2017}) which also come with large uncertainties. 
With the high angular resolution and high sensitivity observations of \textit{AXIS} and \textit{Lynx}, we will be able to accurately map the metal abundance out to 2$r_{200}$ for the first time, heralding a new era in our understanding of the metal enrichment of the ICM and IGM (Fig.\ref{metals}, left). 

\vspace{-4mm}
\section{Key advances in observation capability required}
\vspace{-2mm}

Over the next decade, and into the 2030s, transformative advances in the spectral resolution (\textit{XRISM} \cite{Tashiro2018} and \textit{Athena}) and the grasp (\textit{eROSITA}\cite{Predehl2010,Merloni2012}) of X-ray observatories will be made, allowing us to continue the rapid advancement in the field of cluster outskirts \cite{ettori2013,Walker2019}. 
However all of these missions are fundamentally limited by their large PSFs. An enormous gap in our science capability will remain due to the lack of high spatial resolution, high sensitivity X-ray imaging. Such capability is also vital to allow us to fully exploit improvements in the resolution and sensitivity of SZ observations \citep[][for a review and white papers]{Mroczkowski2019,Mroczkowski2019b,Sehgal2019}.
In order to probe the diffuse gas in the outskirts of galaxy clusters and the cosmic web, an X-ray telescope is required which satisfies 3 key criteria: 

\vspace{-2mm}
\begin{itemize}
\item An effective area to background ratio at least 50$\times$
better than \textit{Chandra's} must be combined with a low and stable particle background
level. The orbits of \textit{Chandra} and \textit{XMM-Newton} are ill suited to this. A low-Earth orbit, 
placing the telescope within the protection of the
Earth's magnetic field, provides the lowest and most stable particle
background. An orbit around L1 or L2 also promises to provide a stable particle background.
\vspace{-2mm}
\item Subarcsecond spatial resolution must be achieved across a large
FoV with low levels of vignetting. This is required to ensure
background AGN in the cosmic X-ray background (CXB) can be resolved and removed
throughout the FoV.  As shown in the top-right panel of Fig.\ref{LSS}, our ability to detect faint diffuse X-ray emission is limited by our ability to resolve out AGN point sources. 
Furthermore, gas clumps associated with infalling galaxies and groups are either too faint to be resolved directly, or impossible to distinguish from background AGN due to low photon counts. A telescope is needed which can resolve these faint gas clumps as extended objects to separate them from background AGN.
\vspace{-2mm}
\item A large FoV with low vignetting is required to allow the large areas
subtended by clusters to be explored within a feasible exposure time. For a
massive ($M_{200}=10^{15}$M$_{\odot}$) cluster at an intermediate redshift of 0.1, the virial radius is around 20 arcmins. To allow the cluster outskirts to be revealed in
their entirety with reasonable exposure times, the size of the imaging detector
therefore needs to be at least 20$\times$20 arcmins, with low vignetting. 
\end{itemize}
\vspace{-2mm}
The imaging detectors on \textit{AXIS/Lynx} fulfill all the criteria
listed above (see also Fig. \ref{metals}, right). With these new instruments, we can measure the thermodynamical properties and chemical composition of the bulk of the ICM up to 2r$_{200}$ (8 times more volume than in the best current studies) and into the filamentary cosmic web.


\pagebreak

\bibliographystyle{unsrturltrunc6}
\bibliography{outskirts_WP}

\end{document}